\documentclass[preprint]{aastex}

\usepackage{textcomp}  
\usepackage{amsmath}    
\usepackage{geometry}

\shorttitle{Steady-State MHD Flow Around a Conducting Sphere}
\shortauthors{Romanelli et al.}

\begin{document}


\title{Steady-State Magnetohydrodynamic Flow Around an Unmagnetized Conducting Sphere}


\author{N. Romanelli, D. G\'omez\altaffilmark{1} and C. Bertucci}
\affil{Group of Astrophysical Flows, Instituto de Astronom\'ia y F\'isica del Espacio, Buenos Aires, Argentina.}
\email{ Corresponding author: N. Romanelli. Tel: +54 1147890179 int: 134 \\ nromanelli@iafe.uba.ar }
 \and
 
\author{M. Delva}
\affil{Space Research Institute, Graz, Austria.}
\email{Magda.Delva@oeaw.ac.at}


\altaffiltext{1}{Departamento de F\'isica, Facultad de Ciencias Exactas y Naturales, Universidad de Buenos Aires, Buenos Aires, Argentina.}


\begin{abstract}

The non-collisional interaction between conducting obstacles and magnetized plasma winds can be found in different scenarios, 
from the interaction
occurring between regions inside galaxy clusters to the interaction between the solar wind and Mars, Venus, active comets or 
even the interaction between 
Titan and the Saturnian's magnetospheric flow. 
 These objects generate, through several current systems, perturbations in the streaming magnetic field 
  leading to its draping around the obstacle's effective conducting surface.
  Recent observational results suggest that several properties associated with the magnetic field draping, such as 
the location of the polarity reversal layer of the induced magnetotail, are affected by variations in the 
conditions of the streaming magnetic field. 
To improve our understanding of these phenomena, we perform a characterization of several magnetic field draping 
signatures by analytically solving an ideal problem in which a perfectly conducting magnetized plasma (with frozen-in 
magnetic field conditions) flows around a spherical body for various orientations of the streaming magnetic field. 
{In particular, we compute the shift of the inverse polarity reversal layer as the orientation of the background magnetic field is changed.}

\end{abstract}


\keywords{Conduction, Magnetohydrodynamics (MHD), Plasmas}



\section{Introduction}
\label{Introduccion}

The interaction between a magnetized plasma flow and a conducting obstacle constitutes a problem of main interest in 
space plasma physics. This phenomenon is observed in a wide range of scenarios, from the solar wind interaction with
unmagnetized planets like Mars or Venus, the solar wind interaction with comets, 
the Saturnian's magnetospheric flow interaction with Titan, to  
the flow of interstellar winds past stars or the interaction between different spatial regions within galaxy clusters.

There are several numerical studies which improved our understanding of the physics involved in
these environments. \cite{dursi08} studied the morphology and the dynamical properties of the streaming magnetic
field perpendicular to the flow moving around a spherical object. In particular, they have shown that these characteristics have important consequences
for galaxy cluster physics because they can suppress hydrodynamic instabilities at the interface of bubbles of active galactic nuclei (AGN). 
In addition, it
provides an explanation to the so-called \textquotedblleft cold
fronts\textquotedblright $\,$ since it allows to sustain strong temperature and density gradients of merging cores that would otherwise
be smoothed out by thermal conduction and diffusion. 
{In the context of our own solar system,} \cite{ma} applied a three 
dimensional multispecies magnetohydrodynamic (MHD) model aimed at studying and 
characterizing the interaction between 
Mars and the Solar Wind. More than ten years before this study, \cite{schmidt} resolved and described the main properties of the solar wind
flow past an active comet by means of a 
global MHD model. Specifically, they studied the deceleration of the solar wind outside the bow shock and
the cometary tail in the downstream region. The 
properties associated with the surrounding environments of these objects show 
interesting similiarities between themselves and with those of Venus and Titan \citep{spreit,bertucci11}. In particular, 
three dimensional MHD simulations characterizing the interaction between Titan and the Saturn's magnetospheric plasma 
can be found in \cite{ledvina}.
Basically, the presence of an unmagnetized body surrounded by an atmosphere (such as Venus, Mars, Titan and the comets) in a
flow of collisionless magnetized plasma (the Solar Wind or the Saturnian magnetospheric plasma) produces 
(through several ionizing mechanisms) a perturbation 
in the plasma and the magnetic field in the region upstream from the object.
These perturbations in the external streaming magnetic field are generated 
by currents originating in  
the differential motion between the external and the planetary or satellite particles.  Because of this,
these environments are called \textquotedblleft induced magnetospheres\textquotedblright, in order to differentiate them from the 
intrinsic magnetospheres where the perturbation is due to the planet's intrinsic magnetic field.
In spite of their differences (atmospheric composition, 
size and upstream magnetic field/flow conditions), these unmagnetized atmospheric objects perturb the streaming
magnetized plasma in a similar fashion, 
thus suggesting that common physical processes might be present. 
In the regions where the collisionless regime holds, the external magnetic field 
frozen into the plasma piles up around the stagnation region of the flow, and drapes around the object as the
flow diverts around the body. This  establishes an induced magnetic tail formed by two lobes 
of opposite magnetic polarity separated by a current sheet \citep{bertucci11}.
{Several properties of the induced magnetospheres depend upon the direction of the 
external magnetic field. In particular, recent observational results around Titan by \cite{simon} suggest that the 
location and shape of the so-called polarity reversal layer (located between both magnetic lobes) are affected by this parameter. The polarity 
reversal layer (hereafter PRL) contains the points where the magnetic field component in the direction of the flow changes sign.
In the case where the external magnetic field is strictly perpendicular to the direction of the flow, 
the induced magnetic tail has a flat structure and the mirror symmetry is maintained. As the background magnetic field departs from being perpendicular 
to the flow, the mirror symmetry breaks down and the polarity reversal layer is shifted.
 }

Previous theoretical studies on these topics have been performed by several 
authors, (e.g. \cite{chacko97,dursi08}). In the first paper, the authors 
derive an analytical 
solution to the problem of a magnetized plasma flowing around a conducting sphere
in steady state conditions 
focusing on two asymptotic regions: very
close and very far from the sphere. They showed that when resistivity 
is included, the magnetic field drapes around the 
sphere, forming a layer of very intense magnetic field with a 
scale size of the order of $O(\eta^{1/2})$, where $\eta$ is the magnetic diffusivity. In the second paper, the authors
follow
an analytical approach to study the physical processes ocurring when an MHD fluid  with
frozen-in magnetic field conditions flows around a sphere and analyze the characteristics of the magnetic field near
the surface of the body. These studies are restricted to the particular case where the angle between the 
background magnetic field and the flow velocity at
infinity is $90$\textdegree.

In the present, we derive for the first time
the analytical solution for the electric and magnetic fields
associated with a perfectly conducting magnetized plasma flowing around a spherical body 
for an arbitrary orientation of the background magnetic field.
{We also identify an additional layer, here called inverse polarity reversal layer (IPRL). 
This layer is characterized by a change of sign in the magnetic field component parallel to the flow which is opposite
to the one in the PRL.
Finally, we characterize the location and shape of the IPRL as a function of the tilt of the background magnetic field.}

The article is structured as follows.
In section (2) we derive the convective electric field related to the interaction between
a MHD plasma flow and an unmagnetized conducting sphere under steady-state conditions.
In section (3) we obtain the magnetic field associated to this problem for any possible angle between the velocity field and the
background magnetic field at infinity.
In section (4) we characterize {the main properties of both polarity reversal layers and briefly discuss the 
changes produced by the inclusion of resistivity effects.}
Finally in section (5) we present a discussion of the results together with our conclusions.

\section{Determination of the electric field}
\label{electric_secion}


Let us consider an MHD flow with frozen-in magnetic field
conditions  whose velocity field  $\boldsymbol{V}(\boldsymbol{r})$ is flowing 
around a sphere of radius $R$ under steady-state conditions. First of all, it is important to notice that this problem 
is inherently three-dimensional: if the external plasma is
not magnetized, the problem still has a symmetry 
about the flow axis; however, when we consider an arbitrary magnetic
field direction frozen to the flow, this symmetry is lost.
It is also relevant to take into account that the frozen-in magnetic field condition is in general justified in 
the enviroments we are interested in, because of the very high magnetic Reynolds numbers that are typical in astrophysical flows.
Therefore, in a first approximation, we determine the electric and magnetic fields neglecting the effects of resistivity.
After this and in order to see the range of validity of this assumption, 
 we  heuristically consider the main changes introduced by resistivity in section \ref{PRLS}.

The equations of ideal MHD with infinite conductivity are the following:

\begin{equation}
\boldsymbol{\nabla} \times (\boldsymbol{V} \times \boldsymbol{B}) = 0 
\label{frozen}
\end{equation}

\begin{equation}
\boldsymbol{\nabla} \cdot \boldsymbol{B} = 0
\label{divergenless}
\end{equation}
where 

\begin{equation}
 \boldsymbol{E}=- \boldsymbol{V} \times \boldsymbol{B}
\label{electricfield}
 \end{equation}
where $\boldsymbol{V}$, $\boldsymbol{B}$ and $\boldsymbol{E}$ are the flow velocity, the magnetic and the electric fields, respectively.

\begin{figure}[!ht]
\begin{center}
\includegraphics[scale=.6]{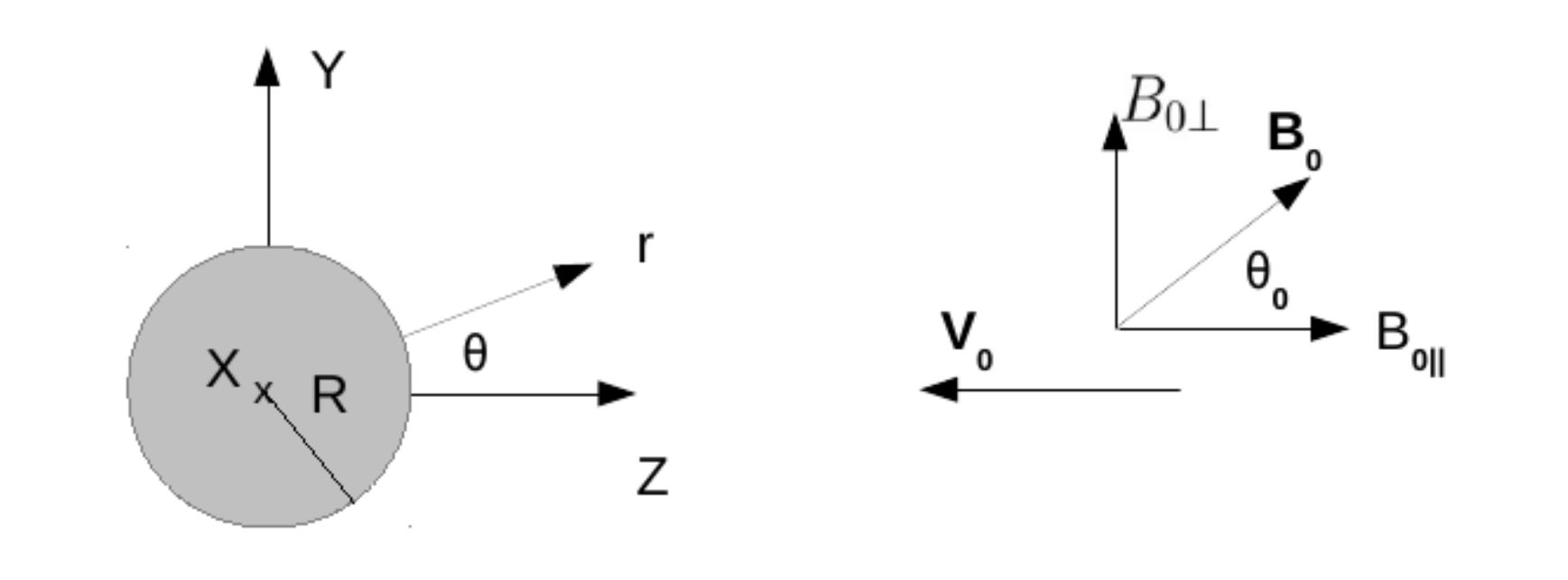}
\caption{Definition of the reference frame. The polar angle $\theta$ is the angle between the +z-axis and a particular position $\boldsymbol{r}$ and it takes
values between 0 and $\pi$. The azimuthal angle $\phi$ is contained in the (x,y) plane and takes values between 0 and $2\pi$ where $\phi=0$ corresponds
to the direction of the x-axis.}
 \label{coordinates}
\end{center}
\end{figure}

We solve this set of equations outside the sphere for a given stationary velocity field which is associated to an ideal and
incompressible flow around the sphere. 
For simplicity, we neglect any change in the flow pattern caused by the back-reaction 
of the magnetic field.
We choose the origin of our coordinate system at the center of the 
sphere and the z-axis being anti-parallel to the fluid velocity at infinity. 
We assume an homogeneous magnetic field $\boldsymbol{B_{0}}$ at infinity  whose component perpendicular to the fluid velocity $\boldsymbol{V_{0}}$  points towards 
the positive y-coordinate axis. Figure \ref{coordinates} shows the adopted reference frame.

Therefore, the flow velocity field, in spherical coordinates results:

\begin{equation}
\boldsymbol{V}=-(1-R^3/r^3) {V_{0}} \, cos(\theta) \,\hat{e}_{r} + (1+R^3/2r^3) {V_{0}} \, sin(\theta) \, \hat{e}_{\theta}
\label{flowvelo}
\end{equation}
where $r$, $\theta$ and ${V_{0}}$ are the radial distance, the polar angle and the intensity of the velocity field at infinity, respectively.
It is also important to notice that this field satisfies the conditions $\boldsymbol{\nabla}\cdot\boldsymbol{V}=0$ and $\boldsymbol{\nabla}\times \boldsymbol{V} = 0$.


We define the impact parameter $p$ of a given line of flow, that is, 
the distance $p$ of a streamline from the z-axis for r $\rightarrow{\infty}$, as

\begin{equation}
 p = \sqrt{1-R^3/r^3} \, \, r\, sin(\theta)
\end{equation}

Equations \ref{frozen} and \ref{electricfield} allow us to express the electric field in terms of 
the electrostatic potential $\Phi$ ($\boldsymbol{E}=-\boldsymbol{\nabla}\Phi$), and therefore

\begin{equation}
 \boldsymbol{V}\cdot\boldsymbol{\nabla}\Phi=0
\label{potential}
 \end{equation}

The boundary conditions associated to equation \ref{potential} are 
that the surface of the sphere is an equipotential surface and 
that at large distances from the object, the electric field goes to an asymptotic uniform value which in our case points along 
the negative x-axis (see equation \ref{electricfield}). Therefore the solution has to fulfill the following boundary condition:

\begin{equation}
\Phi({r\rightarrow \infty})= E_{0} \,r \,sin(\theta) \,cos(\phi) 
\label{boundcond}
\end{equation}
where $\phi$ is the azimuthal angle related to the spherical coordinate system considered in this study.
The solution to equation \ref{potential} consistent with the stated boundary condition (equation \ref{boundcond}) is

\begin{equation}
\Phi=E_{0} \,r \, \sqrt{1-R^3/r^3} \, sin(\theta)\, cos(\phi) = E_{0} \,p \, cos(\phi)
\label{pot}
\end{equation}

Finally, using equation \ref{pot}, the convective electric field in spherical coordinates results:

\begin{equation}
E_{r}= - E_{0} \, \frac{[1+R^3/2r^3]}{\sqrt{1-R^3/r^3}}\,cos(\phi)\, sin(\theta)
\label{Er}
\end{equation}

\begin{equation}
E_{\theta}= -  E_{0} \, \sqrt{1-R^3/r^3}\,cos(\phi)\, cos(\theta)
\label{Etita}
\end{equation}

\begin{equation}
E_{\phi}= E_{0}  \, \sqrt{1-R^3/r^3}\,sin(\phi)
\label{Ephi}
\end{equation}

Equation \ref{electricfield} together with the flow velocity and the electric field, allow us to determine the magnetic field 
components perpendicular to $\boldsymbol{V}$ at each point. However, in order to obtain the total magnetic field we need to take into 
account equations \ref{frozen} and \ref{divergenless} simultaneously.




\section{Determination of the magnetic field}

In order to determine the magnetic field topology for any possible (uniform) background magnetic field orientation, we 
decomposed the problem in two separate contributions: (1) the 
problem in which the angle between the  background 
magnetic field $\boldsymbol{B}_{0}$ and the velocity flow at infinity
$\boldsymbol{V}_{0}$ is 90\textdegree; (2) 
the 
problem in which the angle between $\boldsymbol{B}_{0}$ and $\boldsymbol{V}_{0}$ at infinity is 0\textdegree. 
Because of the linearity of the idealized problem with respect to $\boldsymbol{B}$ (having fixed the velocity field $\boldsymbol{V}$), the general 
solution for any possible angle between both 
fields is simply a linear combination between these two solutions.
Hence, we decompose the magnetic field at infinity as: 

\begin{equation}
\boldsymbol{B}_{0} = \boldsymbol{B}_{0\parallel} + \boldsymbol{B}_{0\perp}
\label{decomposition}
\end{equation}

We first consider the case of a uniform background magnetic field pointing towards the positive $\hat{y}$-coordinate 
system axis $\boldsymbol{B}_{0\perp}$. In spherical coordinates, this field can be written as:

\begin{equation}
(B_{0\perp})_{r}= B_{0\perp} \,  sin(\theta) \,sin(\phi)
\end{equation}

\begin{equation}
(B_{0\perp})_{\theta}= B_{0\perp} \,  cos(\theta) \,sin(\phi)
\end{equation}

\begin{equation}
(B_{0\perp})_{\phi}= B_{0\perp} \,  cos(\phi)
\label{bphibc}
\end{equation}

The spherical components of the differential equations \ref{frozen} and \ref{divergenless} are:

\begin{equation}
 [\boldsymbol{\nabla}\times(\boldsymbol{V}\times\boldsymbol{B})]_{r}=\frac{\partial}{\partial\theta}[sin(\theta) \,(V_{r}\,B_{\theta}-V_{\theta}\,B_{r})]+\frac{\partial}{\partial\phi}(V_{r}\,B_{\phi})=0
 \label{1}
\end{equation}

\begin{equation}
 [\boldsymbol{\nabla}\times(\boldsymbol{V}\times\boldsymbol{B})]_{\theta}=\frac{\partial}{\partial \,r}[r \,(V_{r}\,B_{\theta}-V_{\theta}\,B_{r})]-\frac{1}{sin(\theta)} \, \frac{\partial}{\partial\phi}(V_{\theta}\,B_{\phi})=0
\label{2}
 \end{equation}

\begin{equation}
 [\boldsymbol{\nabla}\times(\boldsymbol{V}\times\boldsymbol{B})]_{\phi}=\frac{\partial}{\partial \,r}(r \,V_{r}\,B_{\phi})+ \, \frac{\partial}{\partial\theta}(V_{\theta}\,B_{\phi})=0
\label{3}
 \end{equation}

\begin{equation}
 \boldsymbol{\nabla}\cdot\boldsymbol{B}=\frac{1}{r^2} \, \frac{\partial}{\partial \,r}(r^2 B_{r})+\frac{1}{r\,sin(\theta)} \, \frac{\partial}{\partial\theta}[sin{(\theta)}\,B_{\theta}]+\frac{1}{r\,sin(\theta)} \, \frac{\partial\,B_{\phi}}{\partial\phi}=0
\label{4}
 \end{equation}

For the flow velocity field $\boldsymbol{V}$ given in equation \ref{flowvelo} and from equation \ref{3} we can derive the differential equation for ${B}_{\phi}$:

\begin{equation}
 \frac{\partial B_{\phi}}{\partial \,r}+\, \frac{V_{\theta}}{r\,V_{r}}\frac{\partial B_{\phi}}{\partial \, \theta}= - \frac{3\,B_{\phi}\,R^3}{2\,r\,(r^3-R^3)}
\label{5}
 \end{equation}

On the other hand, to determine ${B}_{r}$ and ${B}_{\theta}$ we 
multiply equation \ref{1} by $r$ and add equation \ref{2} multiplied by $sin(\theta)$, to obtain

\begin{equation}
 \frac{\partial K}{\partial \,r}+\, \frac{V_{\theta}}{r\,V_{r}}\frac{\partial K}{\partial \, \theta}= 0
\label{6}
\end{equation}
where $K=r\,sin(\theta) \,(V_{r}\,B_{\theta}-V_{\theta}\,B_{r})$.

The resulting linear inhomogeneous first order partial differential equations \ref{5} and \ref{6} can be solved by 
the method of characteristics. 
To this end,  we consider r as a parameter in the characteristic 
equations and express the variables $\theta$ and $\phi$ in terms of r. Therefore, along a streamline:

\begin{equation}
 \frac{d}{dr}=\frac{\partial}{\partial \,r}+\frac{V_{\theta}}{r\,V_{r}}\frac{\partial}{\partial \,\theta}
\end{equation}
since $V_{\phi}=0$ (see Eqn (\ref{flowvelo})).

Along a streamline with impact parameter $p$ at infinity, equation \ref{5} can be rewritten as:

\begin{equation}
 \frac{d B_{\phi}}{d r}= - \frac{3\,B_{\phi}\,R^3}{2\,r\,(r^3-R^3)}
\label{8}
 \end{equation}
while equation \ref{6} becomes:

\begin{equation}
 \frac{d K}{d r}= 0
\label{9}
 \end{equation}

We integrate equation \ref{8} using the boundary condition for the magnetic field at infinity (equation \ref{bphibc}). Therefore:

\begin{equation}
B_{\phi \perp} = \frac{B_{0\perp} cos(\phi)}{\sqrt{1-R^3/r^3}}
\label{Bphi}
\end{equation}
  while from equation \ref{9} we obtain that
 
\begin{equation}
 K=-p\,V_{0}\,B_{0\perp}\,sin(\phi)
\label{K}
 \end{equation}
 
Note that equations \ref{Bphi} and \ref{K} are intimately 
 related to equations \ref{Er}, \ref{Etita} and \ref{Ephi}.
 We can check the results regarding the electric field (obtained in section \ref{electric_secion}) in the following way:
 having determined $B_{\phi }$ and $K$, we calculate:
 
 \begin{equation}
E_{r}=-V_{\theta}\,B_{\phi}=- (1+R^3/2r^3) V_{0} \, sin(\theta)\,\frac{B_{0\perp} cos(\phi)}{\sqrt{1-R^3/r^3}}=-E_{0} \frac{[1+R^3/2r^3]}{\sqrt{1-R^3/r^3}}\,{cos(\phi)}\, sin(\theta)
 \end{equation}

  \begin{equation}
E_{\theta}=V_{r}\,B_{\phi}=-(1-R^3/r^3) V_{0} \, cos(\theta)\,\frac{B_{0\perp} cos(\phi)}{\sqrt{1-R^3/r^3}}=-E_{0} {\sqrt{1-R^3/r^3}}\,{cos(\phi)}\, cos(\theta)
 \end{equation}

   \begin{equation}
E_{\phi}=-(V_{r}\,B_{\theta}-V_{\theta}B_{r})=-\frac{K}{r\,sin(\theta)}=\frac{p\,V_{0}\,B_{0\perp}\,sin(\phi)}{r\,sin(\theta)}=E_{0}  \, \sqrt{1-R^3/r^3}\,sin(\phi)
 \end{equation}
where $E_{0}=B_{0 \perp} V_{0}$.

 Equation \ref{K} couples the magnetic field components $B_{r}$ and $B_{\theta}$.
 Making use of equations \ref{Bphi}, \ref{K} and \ref{4} and applying the method of characteristics along streamlines,
 we obtain the following differential equations for $B_{r}$ and $B_{\theta}$ (see also \cite{dursi08} for a similar derivation):
 
 \begin{equation}
  \frac{d B_{r}}{d r} + [\frac{2}{r}-\frac{2r^3+R^3}{2\,r\,(r^3-R^3)}({1+\frac{1}{cos^2(\theta)}})]B_{r}=-\frac{B_{0\perp}\,sin(\phi)\,sin(\theta)}{r\,\sqrt{1-\frac{R^3}{r^3}}\,cos^2(\theta)}
 \end{equation}

 \begin{equation}
  \frac{d B_{\theta}}{d r} + [\frac{2}{r}-\frac{2r^3+R^3}{2\,r\,(r^3-R^3)}+\frac{9\,r^2\,R^3}{(2r^3+R^3)(r^3-R^3)}]B_{\theta}=\frac{2\,B_{0\perp}\,sin(\phi)(r^3+2R^3)}{r\,\sqrt{1-\frac{R^3}{r^3}}\,cos(\theta)(2r^3+R^3)}
 \end{equation}
 
 The resulting linear inhomogeneous first-order ordinary differential equations are
solved by an integrating factor which is obtained from the homogeneous equations. Finally, 
the magnetic field components $B_{r}$ and $B_{\theta}$ are:

\begin{equation}
B_{r\perp} = \frac{(r^3-R^3)}{r^3} \, cos(\theta) \, \{C_{1} \mp B_{0\perp} sin(\phi)  \int_\infty^r \frac{p \, r^4 \, dr}{(r^3-R^3-p^2\,r)^{3/2} \sqrt{r^3-R^3} }\}
\end{equation}

\begin{equation}
B_{\theta \perp} = \frac{(2 \, r^3 + R^3)}{r^{5/2} \sqrt{r^3-R^3}} \,  \{C_{2} \pm 2\, B_{0\perp} sin(\phi)  \int_\infty^r \frac{r^3 \, (r^3 + 2\,R^3) \, \sqrt{r^3-R^3} \, dr}{(r^3-R^3-p^2\,r)^{1/2}  (2\, r^3 + R^3)^2 }  \}
\end{equation}
  where the upper signs refer to the region $0\leq\theta\leq\pi/2$ and the lower sign to $\pi/2\leq\theta\leq\pi$ and 
  $C_{1}$ and $C_{2}$ are integration constants determined to satisfy the boundary conditions upstream from the object.

To  obtain the magnetic field configuration associated with the more general case 
of an oblique (uniform) background magnetic field at infinity, we now consider the
case of  $\boldsymbol{B}_{0}=\boldsymbol{B}_{0\parallel}$ (therefore $\boldsymbol{B}_{0\perp}=0$).
Note that: $\boldsymbol{\nabla}\cdot\boldsymbol{B}=\boldsymbol{\nabla}\cdot\boldsymbol{V}=0$,  
$\boldsymbol{\nabla}\times(\boldsymbol{V}\times\boldsymbol{B})=0$ and the boundary
conditions (fields at infinity) $\boldsymbol{B_{0}}=\boldsymbol{B}_{0\parallel}$, 
$\boldsymbol{V}=\boldsymbol{V}_{0\parallel}$ are 
fields which point in the direction of the positive/negative z-axis, respectively. Because of this, as long as we are restricted 
to a uniform magnetic field at infinity in the z-direction, the $-\boldsymbol{V}$ and $\boldsymbol{B}$ fields
are characterized by the same topology at any point in space.
Therefore the resulting
magnetic field components in this case are:

\begin{equation}
B_{r \parallel} = (1-R^3/r^3) B_{0 \parallel} \, cos(\theta) 
\end{equation}

\begin{equation}
B_{\theta \parallel} = -(1+R^3/2r^3) B_{0 \parallel} \, sin(\theta)  
\end{equation}

\begin{equation}
B_{\phi \parallel} = 0 
\end{equation}

As we previously pointed out, 
the general solution for any possible orientation of the magnetic field at infinity
is simply a linear combination between the two solutions proposed in equation \ref{decomposition}.
If $\theta_{0}$ is the angle between the background magnetic field and the z-axis (see figure \ref{coordinates}), then
$|B_{0 \parallel}|=B_{0} \,cos(\theta_{0})$ and $|B_{0 \perp}|=B_{0} \,sin(\theta_{0})$.

\begin{figure}[!ht]
\begin{center}
\includegraphics[scale=.72]{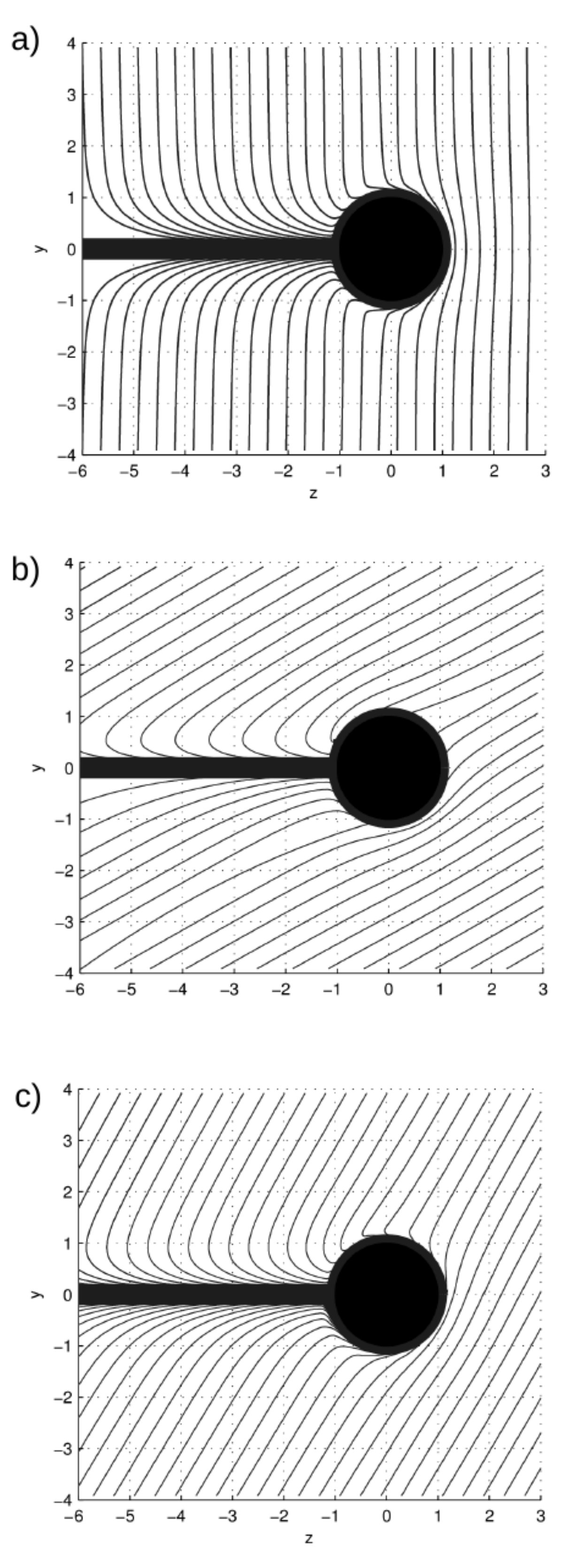}
\caption{\scriptsize{ Panel a) Predicted magnetic field lines around the sphere for $\boldsymbol{B}_{0} \perp \boldsymbol{V}_{0}$.
Panel b) Predicted magnetic field lines around the sphere for $\theta_{0}=30$\textdegree.
 Panel c) Predicted magnetic field lines around the sphere for $\theta_{0}=60$\textdegree.}
}
 \label{fig90}
\end{center}
\end{figure}

Figure \ref{fig90}a) shows the magnetic field lines around 
the spherical obstacle on the y-z plane, for the 
case where $\boldsymbol{B}_{0} \perp \boldsymbol{V}_{0}$ (hence $\boldsymbol{B}_{0}$ parallel to the y-axis).
We can observe the following properties: 

\begin{itemize}
 \item There is an increase of the magnetic field intensity at the surface of the sphere 
 and also in the downstream region, where the largest values 
are reached in the surroundings of the $Y = 0$ plane.
 \item {The magnetic field structure consists of two mirror symmetric magnetic 
hemispheres separated by a flat polarity reversal layer located on the plane $Y=0$ 
(the thick black bar in figure \ref{fig90}).} This 
change of the polarity of the flow aligned-component of the magnetic field 
as well as the increase of the magnetic field strength close to the surface of the sphere are main features 
associated to a draping configuration.
\end{itemize}

Figures \ref{fig90}b) and \ref{fig90}c) show the magnetic field lines around the spherical obstacle (on the y-z plane) for 
two orientations of the background magnetic field ($\theta_{0}=30$\textdegree and $\theta_{0}=60$\textdegree, respectively).
{For cases of non-perpendicular magnetic field and flow direction, the
above mentioned symmetry is broken, giving rise to an inverse polarity 
reversal layer (IPRL). The IPRL observed in figures  \ref{fig90}b) and \ref{fig90}c) corresponds to 
the set of points where $B_{z}=0 \,$ right above the PRL. Therefore, in these cases there are two layers where  the 
flow aligned-component of the magnetic field reverses its polarity. One of these layers, 
the PRL is located in the $Y=0$ plane, just as it is observed in the strictly perpendicular $\boldsymbol{B}_{0 \perp}$case. 
In constrast to this, the location and shape of the IPRL varies with $\theta_0$. In the next section, we perform a more detailed analysis of the properties of 
each of these layers.
}

\begin{figure}[!ht]
\begin{center}
\includegraphics[scale=.65]{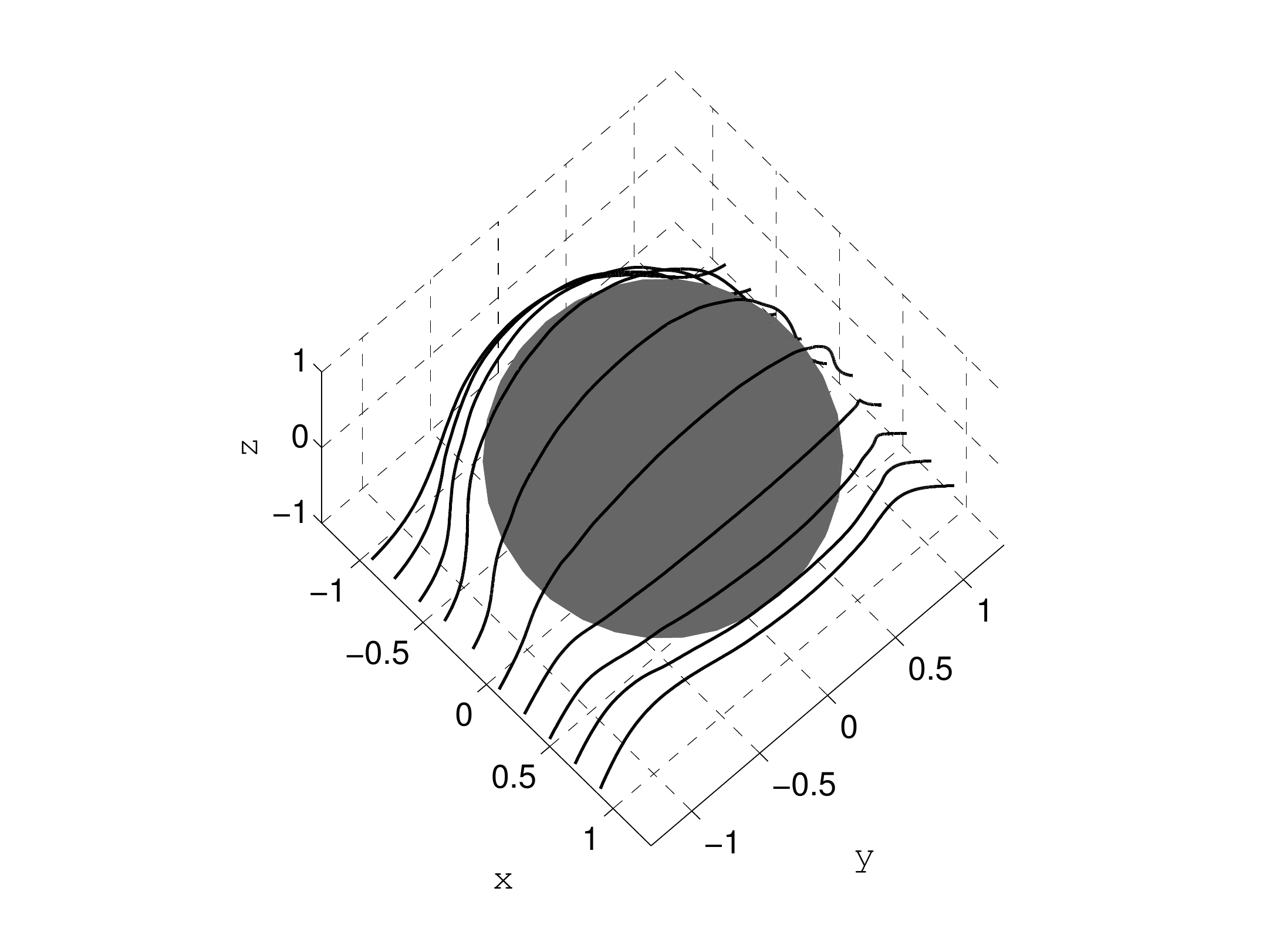}
\caption{Magnetic field lines in the vicinity of the spherical obstacle associated to different equipotential surfaces.}
 \label{fig3d}
\end{center}
\end{figure}

Finally, figure \ref{fig3d} shows magnetic field lines in the vicinity of the spherical obstacle
associated with different equipotential surfaces. Note that because of $\boldsymbol{B} \cdot \boldsymbol{\nabla} \Phi =0$, each magnetic field line is 
contained in a surface in which the electrostatic potential $\Phi$ (equation \ref{pot}) is constant (see \cite{gombosi}).

\begin{figure}[!ht]
\begin{center}
\includegraphics[scale=0.5]{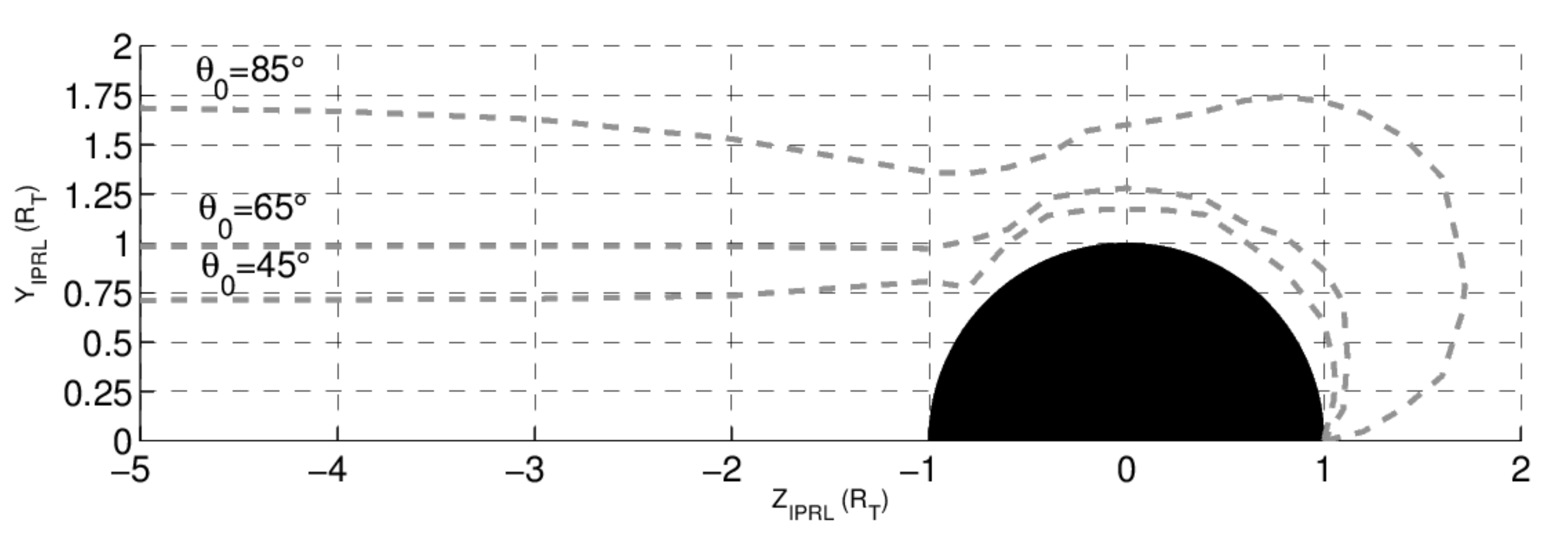}
\caption{Location of the inverse polarity reversal layer (IPRL) in the y-z plane 
for different angles between the background magnetic field and the flow direction. 
 The dashed lines indicate the location for $\theta_0$= 45\textdegree, 65\textdegree $\,$ and 85\textdegree.
}

 \label{IPRL}
\end{center}
\end{figure}

\begin{figure}[!ht]
\begin{center}
\includegraphics[scale=.7]{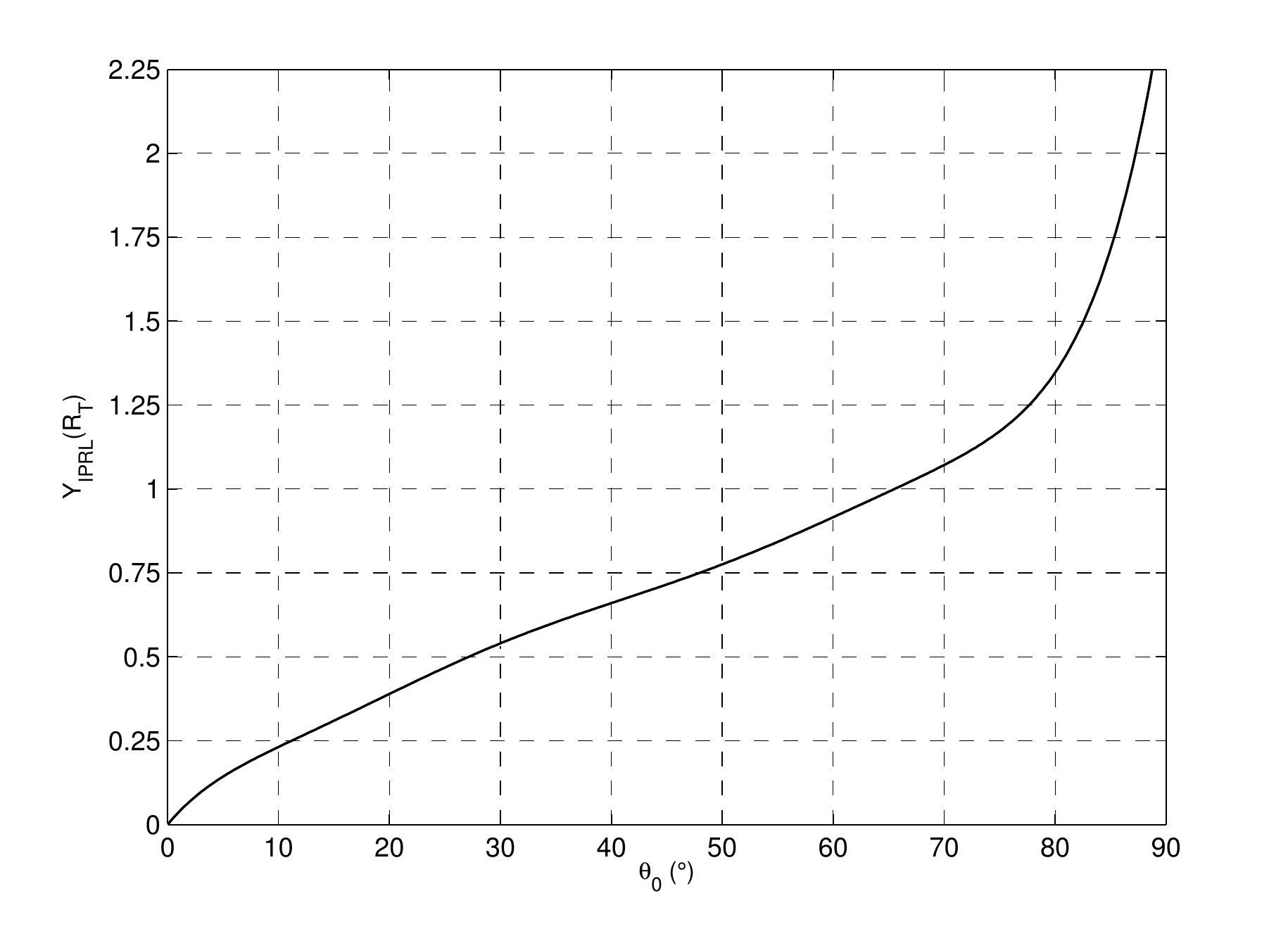}
\caption{Distance of the IPRL layer from the $Y=0$ plane, at Z=$- 3 R_T$,  as function of $\theta_0$.}
 \label{Asymptotic}
\end{center}
\end{figure}

{
\section{Polarity reversal layers}
\label{PRLS}
All previous results were obtained under the assumption that resistivity effects were negligible.
While the resistivity effects are really important in the surroundings of the PRL, they are not expected to affect the location and shape 
of the IPRL significantly. This difference is due to the fact that while the intensity of the magnetic field is infinite at
both sides of the PRL, in the proximity of the IPRL it remains finite and varies rather smoothly.
}
{
\subsection{Polarity reversal layer}
\cite{chacko97} showed that the divergent behavior of the magnetic field in certain locations
is due to the absence of resistivity effects, which therefore need to be taken into account.
They found that the electrostatic potential remains unaffected only at length scales such 
that  $p/R> O(R_{m}^{-1/4})$, where $R_{m}$ is the magnetic Reynolds number.   Moreover, they identified a 
 boundary layer totally contained in the tail region inside of which the resistivity effects are relevant. 
As shown in section (3), this is  the case for the PRL since it is always located in the $Y=0$ plane.
The reader interested in the analytical computation of the magnetic field in this region (with the inclusion of diffusivity effects) 
is referred to \cite{chacko97}. 
 Cassini observations at Titan (see \cite{simon}) 
suggest an observational correlation between $\theta_0$ and the vertical
displacement of the PRL. 
In most cases, it was found that the PRL displacement becomes smaller for larger $\theta_0$.
\subsection{Inverse polarity reversal layer}
Since the resistivity effects are not essential in the region near the IPRL, we determine its position and shape
 for different values of $\theta_0$, using the theoretical approach presented in the previous section.
Figure \ref{IPRL} shows the location and geometry of the IPRL in the Y-Z plane for $\theta_0 =$ 45\textdegree, 65\textdegree $\,$ and 85\textdegree. 
For small values of the z-coordinate (i.e. z such that $-1 R_T<Z_{IPRL}<2 R_T$) 
the IPRL is well draped around the obstacle. For positions further downstream (z $< - 3 R_T$) the IPRL becomes aligned with the flow.
Additionally, in the downstream part with flow-aligned IPRL, this layer
is clearly shifted to larger distances from the  $Y=0$ plane for larger values of the angle $\theta_0$.
Figure \ref{Asymptotic} shows the y-coordinate corresponding to the location
of the IPRL in the downstream region (in particular, for $Z_{IPRL}=-3 R_T$) 
for $\theta_0$ in $[$0\textdegree, 90\textdegree$]$. As it can be seen, the tendency mentioned above
(larger shift for larger $\theta_0$) is observed for all $\theta_0$ angles.
}


\section{Discussion and Conclusions}

We calculated the analytical solution for both the electric and the magnetic field
associated to an ideal plasma flowing around a spherical obstacle for an arbitrary orientation of background magnetic field.
The motivation of this article follows from observations in different spatial enviroments that show, in spite of their differences, significant 
similarties in the electric and magnetic field topologies. Among other examples, we find the interaction between the solar wind and planetary 
plasmas and atmospheric non-magnetized objects; the interaction between regions with different temperature or densities
within galaxy clusters and even in the coupling between 
interestellar winds and stars.

Based on the high magnetic Reynolds numbers associated with these enviroments, we first obtained the electric potential under 
frozen-in magnetic field conditions. Under these conditions,  the electric potential does not vary along 
the streamlines of the flow, as it becomes apparent from equation \ref{potential}. By making use 
of this property and the corresponding boundary condition we derived $\Phi_{ideal}$ and then 
we calculated the corresponding electric field components shown in equations \ref{Er}, \ref{Etita} and \ref{Ephi}.
At this point, it is important to realize that the electric field remains the same regardless of the orientation of the background 
magnetic field as long as $\boldsymbol{B}_{0 \perp}$ does not change. 

To calculate the total magnetic field we needed to consider equations \ref{frozen} and \ref{divergenless} simultaneously.
Therefore, we integrated partial diffential equations for each magnetic field component using the method of characteristics.
The results presented in figure \ref{fig90}a) show a classical draping configuration where the magnetic field 
frozen into the flow piles up at the stagnation point and drapes around the sphere while the
flow is diverted around the body. This generates an induced magnetic tail composed of two lobes 
of opposite magnetic polarity separated by a polarity reversal layer located at the $Y=0$ plane.
 The magnetic field strength in the surroundings of 
 this layer and in the surface of the sphere is infinitely large. 
 Figures \ref{fig90}b) and \ref{fig90}c) show the magnetic field topologies 
 when the background magnetic field at infinity is
 tilted with respect to the z-axis by 30\textdegree $\,$and 60\textdegree, respectively.
 Comparing these figures we notice similarties and differencies 
 with the strictly perpendicular case: among the similarities we find 
 the increment of the magnetic field strength 
 around the obstacle and in the surroundings of the $Y=0$ plane.
 We also identify a corresponding polarity 
 reversal layer at this plane
 and the induced tail formed by two lobes of opposite polarity. 
  However, a significant difference is found: a second
 polarity reversal layer (the IPRL) exists
  in the downstream region and its location changes in response 
  to variations in the background magnetic field orientation.
  The $\boldsymbol{B}_{0 \parallel}$ component gives rise to this layer since,
  at these locations, the magnetic field contribution 
  associated to the $\boldsymbol{B}_{0 \parallel}$ component is cancelled by the contribution of the field related to
  the $\boldsymbol{B}_{0 \perp}$ component. This layer is always present since the field associated to $\boldsymbol{B}_{0 \parallel}$
  does not change the sign of its component along the z-axis at any point. Therefore, it is always
  cancelled by one of the two lobes produced by the perpendicular component of $\boldsymbol{B}_{0}$.
  The results presented in figures \ref{IPRL} and \ref{Asymptotic} show that the larger $\theta_0$, the larger 
  the distance between the IPRL and the $Y=0$ plane.

 {Moreover, the infinitely large magnetic field strength close to the sphere and in the tail region
  (derived for the non-resistive case) are smoothed out by the inclusion of  
 resistivity effects (see \cite{chacko97}), giving rise to the formation of a magnetic boundary layer (in regions where $p/R< O(R_{m}^{-1/4})$). 
 These effects are relevant to determine the location and shape of the PRL, but they are not essential to characterize the IPRL.}
 
 It is interesting to analyze these results in the context of the exchange of 
energy between the external flow and the obstacle, as the mapping of the convective electric
field into the ionospheres of solar system planets is usually invoked to be the driver of the 
planetary plasma dynamics \citep{gombosi}.
As a result of the convection of the external magnetic field, the convective
electric field is responsible for the onset of currents on the obstacle's conducting surface. 
In the absence of dissipation effects, the sphere will act as a simple mean to accumulate magnetic
flux upstream and around the magnetic tail neutral sheet. The inclusion of a finite conductivity in
a region adjacent to the sphere will lead to a resistive heating at the expense of the magnetic energy 
in that region. This heating is then the result of the action of the convective electric field onto the 
sphere in the presence of dissipative effects. 

 In summary, in this study 
 we find that the interaction between a spherical conductor and an ideal magnetized plasma flow
 generates perturbations in the background magnetic field that are consistent with a classical draping configuration.
 Moreover, we find that the draping configuration downstream from the obstacle becomes asymmetric 
 when $\boldsymbol{B}_{0}$ and $\boldsymbol{V}_{0}$ are not perpendicular and that there is a clear correlation between the direction of the external magnetic field
 and the location and geometry of the IPRL.



\acknowledgments
\section*{Acknowledgments}
N.R. is supported by a PhD fellowship from CONICET. We acknowledge the funding from ANPCyT at IAFE through 
grant PICT 0454/2011.

\clearpage

\clearpage

\addtolength{\topmargin}{-.75in}

\clearpage
\addtolength{\topmargin}{.75in}

\clearpage

\clearpage

\clearpage





\end{document}